**Title**: To better understand realized ecosystem services: An integrated analysis framework of supply, demand, flow and use

**Short Title**: A framework for quantifying realized ecosystem services

**Authors**: Shuyao Wu[1,2], Kai-Di Liu[1,2], Wentao Zhang[1,2], Yuehan Dou[3], Yuqing Chen[4], Delong Li[5*]

[1] Center for Yellow River Ecosystem Products, Shandong University, Qingdao, Shandong 266237, China

[2] Qingdao Institute of Humanities and Social Sciences, Shandong University, Qingdao, Shandong 266237, China

[3] Department of Health and Environmental Sciences, Xi'an Jiaotong-Liverpool University, Suzhou, Jiangsu 215123, China

[4] Fenner School of Environment and Society, Australian National University, Canberra ACT 2601, Australia

[5] Institute of Geographic Sciences and Natural Resources Research, Chinese Academy of Science, Beijing, 100101, China

**\* Corresponding to**:
lidelong@igsnrr.ac.cn (Delong Li)

**Email addresses**:
wushuyao@email.sdu.edu.cn (Shuyao Wu)
liukaidi@sdu.edu.cn (Kai-di Liu)
zhangwt@sdu.edu.cn (Wentao Zhang)
Yuehan.Dou@xjtlu.edu.cn (Yuehan Dou)
yuqing.chen1@anu.edu.au (Yuqing Chen)
lidelong@igsnrr.ac.cn (Delong Li)

**Abstract**

Realized ecosystem services (ES) are the actual use of ES by societies, which is more directly linked to human well-being than potential ES. However, there is a lack of a general analysis framework to understand how much ES was realized. In this study, we first proposed a Supply-Demand-Flow-Use (SDFU) framework that integrates the supply, demand, flow, and use of ES and differentiates these concepts into different aspects (e.g., potential vs. actual ES demand, export and import flows of supply, etc.). Then, we applied the framework to three examples of ES that can be found in typical urban green parks (i.e., wild berry supply, pollination, and recreation). We showed how the framework could assess the actual use of ES and identify the supply-limited, demand-limited, and supply-demand-balanced types of realized ES. We also discussed the scaling features, temporal dynamics, and spatial characteristics of realized ES, as well as some critical questions for future studies. Although facing challenges, we believe that the applications of the SDFU framework can provide a systematic way to accurately assess the actual use of ES and better inform management and policy-making for sustainable use of nature's benefits. Therefore, we hope that our study will stimulate more research on realized ES and contribute to a deeper understanding of their roles in enhancing human well-being.

## 1. Introduction

Ecosystem services (ES), which refer to the benefits people obtain from nature, serve as a perfect bridge connecting nature and social systems (Costanza et al., 1997). These services are derived from ecosystem structure and processes and finally contribute to the improvement of human well-being, which is known as the ES cascade (Haines-Young and Potschin, 2010). Due to the wide applicability of the concept, there have been many attempts to incorporate ES into various policies and plans to improve human well-being (Wong et al., 2015). For example, in the recently issued white paper "China's Green Development in the New Era" by China's State Council Information Office, the Chinese government sets the goal of providing "more quality eco-environmental goods to help people feel happier, more satisfied, and more secure in a beautiful environment (https://english.www.gov.cn/archive/whitepaper/202301/19/content_WS63c8c053c6d0a757729e5db7.html). Compared to the ecosystem's potential to provide ecosystem services, the realized ES (i.e., the actual use of ES, used interchangeably in the study) is more directly linked to human well-being and matters more to policy-makers (Jones et al., 2016; Aziz, 2023). Therefore, understanding the amount of realized ES should be the focus of policies that aim to improve human well-being by conserving and sustainably using nature.

Over the more than twenty years of study of ecosystem services, many concepts have been proposed and studied, such as the supply, demand, and flows of ES (Costanza et al., 2017; Peng et al., 2023). These concepts represent different aspects of the ES cascade and are paramount to understanding the actual use of ES. However, none of the concepts are equivalent to realized ES. Firstly, not all supply or demand can be translated into actual use. For instance, Rasmussen et al. (2016) used a combination of four complementary methods to assess the actual use of provisioning services and found that people's actual use of ecosystem services from agricultural fields differs from the service availability; Aziz and Shah (2019) generated the combined spatial distribution and use intensity of a bundle of six ES in Southern Ontario, Canada, and found that the value of realized ES was only about 50% of the value of potential ES; Brauman et al. (2020) used a framework that distinguishes between potential and realized nature's contributions, environmental conditions, and impacts of changes in nature on the quality of life, and found that most of the 18 assessed nature's contributions have declined globally. Moreover, although some studies regard the ES flows as the realized ES (Villamagna et al., 2013; Burkhard et al., 2014), losses in ES during the flow process are also possible, such as visual blight in the aesthetic experience, release of stored carbon due to disturbances, evaporation of clean water before consumption, etc. (Bagstad et al., 2013; Wei et al., 2017). Therefore, the realized ES is influenced by, but different from all of these concepts but can be the key to integrating them.

Despite all of these efforts to evaluate the realized ES, there is still a lack of a general analysis framework to understand how much ES was realized or actually used in most studies (about 300 studies found in the 2022 Web of Science Core Collection using realized OR actual use ecosystem service as topic keywords vs. over 9000 studies using ecosystem service as the keyword). Without such knowledge, it will be challenging to assess the actual contributions of nature to society accurately and form sustainable ecosystem management plans or policies that can also improve human well-being. To narrow this knowledge gap, in this study, we first proposed and explained an analysis framework of realized ES that integrates the supply, demand, flow, and use of ES. Then, we demonstrated its use with examples of three ES that can be found in typical urban green parks and finally discussed several important characteristics and critical questions related to the application of the framework.

## 2. The Supply-Demand-Flow-Use Framework

The proposed Supply-Demand-Flow-Use (SDFU) Framework to analyze realized ES is depicted in Fig. 1. It can be generally divided into four main sections, which are: 1) the delineation of Service Supply Areas (SSA), Service Demand Areas (SDA) and flows; 2) the actual supply of ES evaluation; 3) the actual demand of ES evaluation; and 4) the actual use of ES evaluation and analysis. After completing the analysis for each ES, analyses of multiple realized ES can also be combined and evaluated. In the following sections, we will explain the purpose, logic, and application of each section. A complete definition of the glossary used in the framework can be found in Table 1.

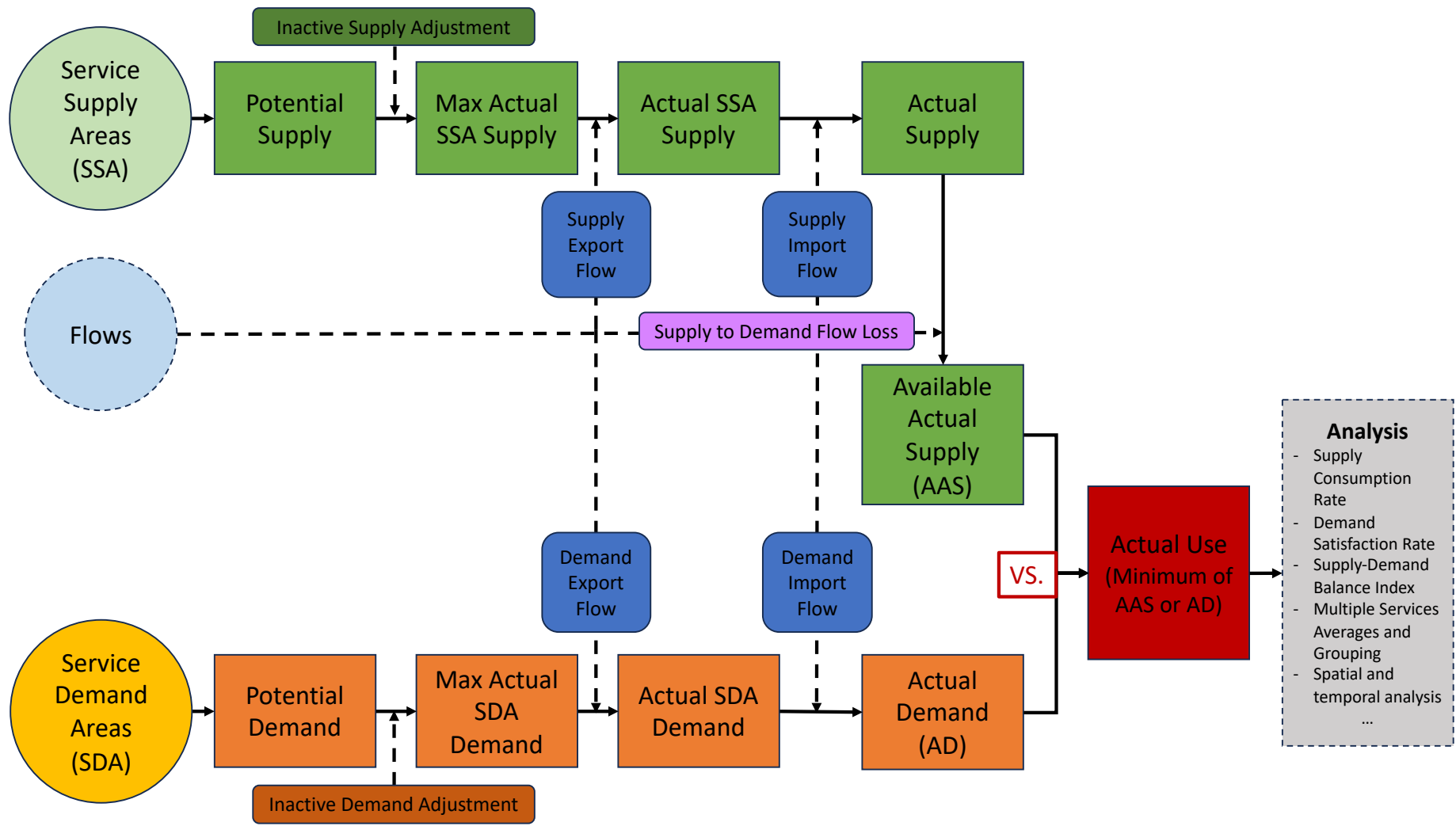

**Fig. 1.** The proposed Supply-Demand-Flow-Use Framework for assessing the realized or actual use of ecosystem services.

**Table 1**

Definitions of the terms used in the Supply-Demand-Flow-Use Framework.

| Variables | Definitions |
| --- | --- |
| *Realized Ecosystem Services* | The amount of potential ecosystem service that is actually used by the service beneficiaries |
| *Service Supply Areas (SSA)* | The spatial units with ecosystem service supply capacity that is under evaluation, also known as Service Provisioning Areas |
| *Service Demand Areas (SDA)* | The spatial units in which ecosystem services are most needed from the *SSA* during evaluation, also known as Service Benefiting Areas |
| *Potential Supply* | The theoretical maximum capacity of ecosystem service supply in *SSA* |
| *Max Actual SSA Supply* | The total amount of activated ecosystem services supply capacity in *SSA* when being evaluated |
| *Actual SSA Supply* | The difference between *Max Actual SSA Supply* and *Supply Export Flow* |
| *Supply Adjustment* | The amount of inactive ecosystem service supply, which is determined by both the internal characteristics of the service providers and the external conditions |
| *Actual Supply* | The addition of *Actual SSA Supply* and *Supply Import Flow* |
| *Potential Demand* | The theoretical maximum amount of ecosystem service demand in *SDA* |
| *Max Actual SDA Demand* | The total amount of activated ecosystem service demand potential in *SDA* when being evaluated |
| *Demand Adjustment* | The amount of inactive ecosystem service demand, which is determined by both the internal characteristics of the service demander and the external conditions |
| *Actual SDA Demand* | The difference between *Max Actual SDA Demand* and *Demand Export Flow* |
| *Actual Demand* | The addition of *Actual SDA Demand* and *Demand Import Flow* |
| *Supply Export Flow* | The amount of ecosystem services supply originated within *SSA*, exited *SSA* through spatial flows, and was consumed outside *SDA* |
| *Supply Import Flow* | The amount of ecosystem services supply originated outside *SSA*, entered SSA through spatial flows, and was consumed within *SDA* |
| *Demand Export Flow* | The amount of ecosystem services demand originated within *SDA,* exited SDA through spatial flows, and had demand satisfied outside *SSA* |
| *Demand Import Flow* | The amount of ecosystem services demand originated outside *SDA*, entered SDA through spatial flows, and had demand satisfied within SSA |
| *Service to Demand Flow* | The spatial flow process of ecosystem services supply from *SSA* to *SDA* |
| *Available Actual Supply* | The difference between *Actual Supply* and the loss of ecosystem service supply during *Service to Demand Flow* |
| *Actual Use* | the minimum of *Available Actual Supply* or *Actual Demand* |
| *Supply Consumption Rate* | The ratio of *Actual Use* to *Available Actual Supply* |
| *Demand Satisfaction Rate* | The ratio of *Actual Use* to *Actual Demand* |
| *Supply-Demand Balance Index* | The difference between *Supply Consumption Rate* and *Demand Satisfaction Rate* |

2.1 SSA, SDA and Flows Delineation

Similar to some existing integrated analysis frameworks of ES supply, demand, and flows (e.g., Serna-Chavez et al., 2014; Schröter et al., 2018; Wang et al., 2023), the first step of the

SDFU framework is to delineate the areas of ecosystem service supply, demand, and flow under evaluation (Fig. 1). We refer to the spatial units with ecosystem service supply capacity that are under evaluation as the Service Supply Areas (SSA), which are also called as the service provisioning or providing areas in other studies (Serna-Chavez et al., 2014; Li and Wang, 2023). On the other hand, we refer to the spatial units in which ecosystem services are needed during evaluation as the Service Demand Areas (SDA), which are also known as the service benefiting areas. For the ES flows, we delineated a total of five different types of flows. The first type is the spatial flow of ES from supply to demand (i.e., supply-to-demand flow), which refers to the spatial flow process of ecosystem services supply from SSA to SDA (Bagstad et al., 2013; Kleemann et al., 2020). This supply-to-demand flow can be both intraregional and interregional, depending on the degree of overlap between the SSA and SDA (Wang et al., 2022).

In addition to the supply-to-demand flow, we also considered four additional types of flows, which are the export and import flows of ES supply and demand (Table 1), to delineate the cross-regional diffusion effects of ES flows (United Nations, 2021; Li and Wang, 2023; Liu, 2023). The reason for such differentiation is that there will be a range of situations in which, for example, the supply of ecosystem services will be used by beneficiaries that are not residents of the SDA (i.e., demand import; e.g., visitors) and also cases where beneficiaries use ecosystem services that originated outside the SSA (i.e., supply import; e.g., food imports) (United Nations, 2021). In such cases, to accurately assess the actual use of ES, export and import flows of both ES supply and demand should be identified for further analysis.

*2.2 Actual Supply of ES Evaluation*

The second step of assessing realized ES involves the evaluation of ES supply in the SSA. In most previous studies defining the ES supply, the supply of ES was not further differentiated or only distinguished the potential and actual supply in a general way (Liu et al., 2023; Zeng et al., 2023). However, to accurately assess the real amount of ES supply in SSA, we believe it is necessary to further distinguish different aspects of the ES supply, which can include *Potential Supply*, *Max Actual SSA Supply*, *Actual SSA Supply* and *Actual Supply* (Fig. 2A). The *Potential Supply* refers to the maximum theoretical capacity of ecosystem service supply in SSA (Table 1). However, not all of the capacity will be activated or able to be translated into actual supply during the evaluation period, and thus adjustments are required to obtain the actual activated amount of ES supply in the SSA (Aziz and Shah, 2019; Liu et al., 2023).

Supply adjustment aims to subtract the amount of inactive ecosystem service supply, which is determined by both the internal characteristics of the service providers (e.g., species composition, ecosystem structure, etc.) and the external conditions (e.g., climate, infrastructure, etc.). Taking the pest control service as an example, although the total population of the pest predators (i.e., the *Potential Supply*) in the SSA can potentially provide this service, only the active (e.g., healthy and mature) proportion of the pest predator population in the SSA (i.e., the *Max Actual SSA Supply*) during the evaluation is really capable of providing such a service and thus could be considered the maximum actual supply of the service in the SSA (Dangles and Casas, 2019). External disturbances, such as drought, flooding, warming, etc., can also further decrease the amount of activated service supply potential (Kremen and Chaplin-Kramer, 2007).

Once we obtain the maximum possible actual supply of ES in the SSA, the next step is the consideration of the export and import flows of ES supply (Fig. 2A). The export/import supply flows of ES refer to the amount of ecosystem services supply originated within/outside SSA, exited/entered SSA through spatial flows, and was consumed outside/inside SDA during evaluation. The export flows need to be subtracted from the *Max Actual SSA Supply* to obtain the *Actual SSA Supply*. Then, the import flows of ES supply need to be added to the *Actual SSA Supply* to obtain the *Actual Supply.* In the example of pest control service, the export and import flows of ES supply can represent the pest predators moving out and into SSA to predate, respectively. Only the actual amount of pest predators in the SSA (to be more specific, the number of pests that these predators can predate) during evaluation can be considered the *Actual Supply* of pest control service.

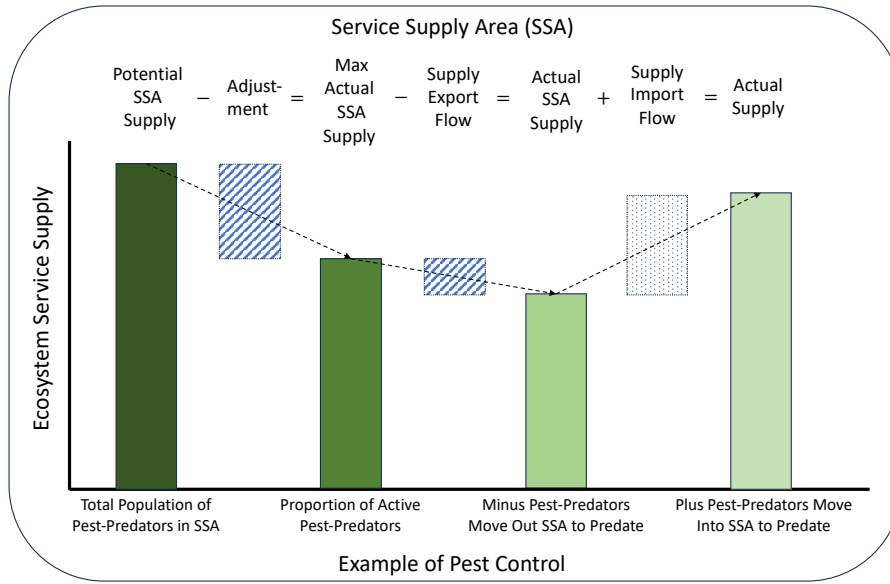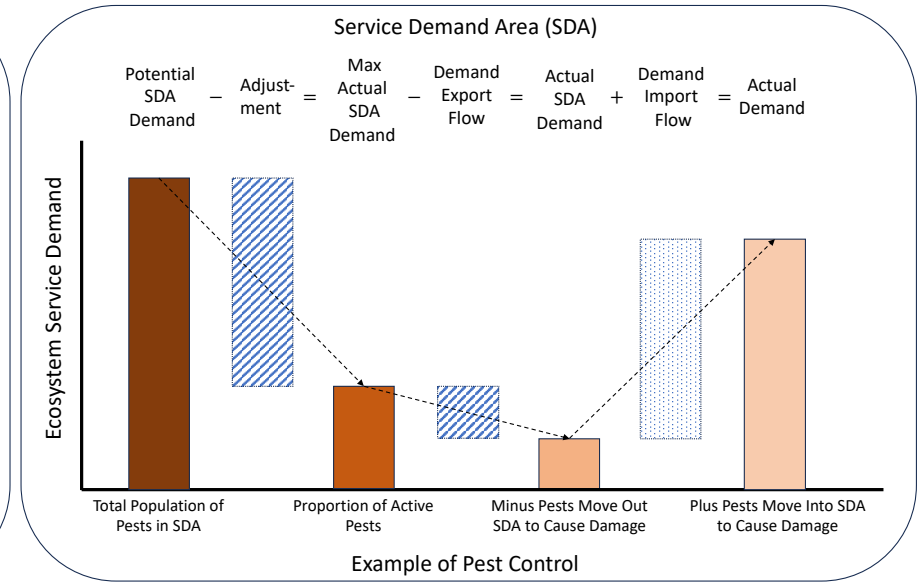

**Fig. 2.** Illustrative evaluation of actual supply and actual demand of ecosystem services with the pest control service as an example.

*2.3 Actual Demand of ES Evaluation*

The next step is the evaluation of ES demand in the SDA. Similar to the ES supply, the demand for ES should also be further differentiated into four aspects (i.e., *Potential Demand*, *Max Actual SDA Demand*, *Actual SDA Demand* and *Actual Demand*) to accurately assess the real amount of demand for ES (Fig. 2B). The potential demand represents the theoretical maximum amount of ecosystem service demand in SDA. Again, adjustments are also needed to turn the theoretical amount into the actual activated amount of demand in the SDA.

Demand adjustment is the amount of inactive ecosystem service demand that needs subtraction, which depends on both the internal characteristics of the service demander (e.g., age structure, personal preferences, etc.) and the external circumstances (e.g., management practice, technology, etc.). Still taking the pest control service as an example, only the active (e.g., healthy and mature) proportion (i.e., the *Max Actual SDA Demand*) of the total pest population in the SDA (i.e., the *Potential Demand*) is really capable of causing damage and thus should be considered the maximum actual demand of the service in the SDA during the evaluation. External management practices such as pesticide usage might also reduce the actual demand for pest predators (Vandermeer et al., 2010).

The export and import flows of ES demand are what we should consider next to obtain the *Actual Demand* for ES (Fig. 2B). The export/import demand flows of ES refer to the amount of ecosystem services demand originated within/outside SDA, exited/entered SDA through spatial flows, and had demand satisfied outside/inside SSA during evaluation. Again, in the example of pest control service, the export and import flows of ES demand can represent the pests moving out and into SDA to cause damage, respectively. Only the number of pests actually causing damage in the SDA during evaluation can be considered the *Actual Demand* for pest control service.

*2.4 Actual Use of ES Evaluation and Analysis*

Before assessing the *Actual Use* of ES, the losses during supply-to-demand flows need to be considered to obtain the *Available Actual Supply* (AAS) of ES (Fig. 3). Some examples of such losses include visual blight in the aesthetic experience, release of stored carbon due to disturbances, evaporation of clean water before consumption, etc. (Bagstad et al., 2013; Wei et al., 2017). But it is worth noting that not all services have losses during the supply-to-demand flow process (e.g., in situ ES such as recreation), and it is possible to have the ES supply replenished during the process (e.g., precipitation to increase water supply in a river flowing from upstream SSA to downstream SDA). However, if the replenishment of the ES supply occurred outside of the SSA under evaluation, then the extra supply should not be accounted for as any form of ES supply in this case since it did not directly originate from the SSA. A careful distinction of such out-of-SSA supply should be made when measuring the supply-to-demand flow data to avoid overestimation of AAS.

Once the AAS is obtained, the *Actual Use* of ES can be assessed by comparing the AAS and *Actual Demand* (AD). A total of three scenarios will be possible after the comparison, namely: 1) Supply is greater than Demand, 2) Supply equals Demand, and 3) Supply is less than Demand. The *Actual Use* of ES is constrained by either the minimum supply or demand quantity of ES; in other words, it equals the lower value of AAS or AD (La Notte et al., 2019). The three scenarios of actual use represent three corresponding types of realized ES, which are supply-limited, supply-demand (S-D) balanced, and demand-limited ES (Fig. 3).

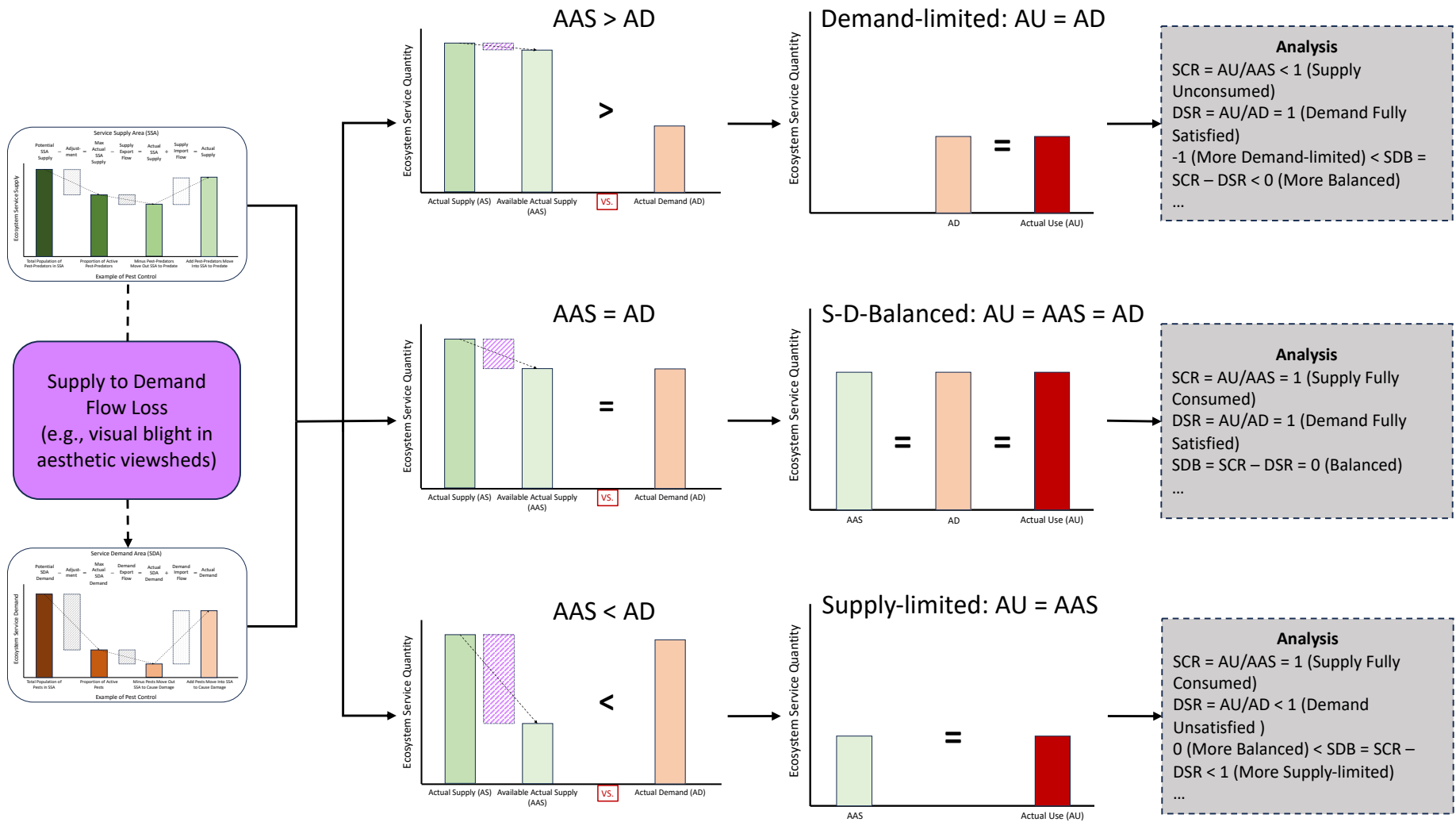

Fig. 3. Illustrative evaluation and analysis of realized ecosystem services under three scenarios (i.e., Demand-limited, Supply-Demand-Balanced, and Supply-limited). SCR: Supply Consumption Rate; DSR: Demand Satisfaction Rate; SDB: Supply-Demand Balance Index.

For analyzing the characteristics of different scenarios of ES realization, multiple coefficients can be developed (Fig. 3). The three most important ones we propose are the Supply Consumption Rate (SCR, the ratio of *Actual Supply* to *Available Actual Supply*, 0<SCR≤1), the Demand Satisfaction Rate (DSR, the ratio of *Actual Supply* to *Actual Demand*, 0<SCR≤1) and the Supply-Demand Balance Index (SDB, the differences between SCR and DSR, -1<SDB<1). The higher the values of SCR and DSR, the more supply and demand of ES are consumed and satisfied, respectively. For the supply-limited and demand-limited types of realized ES, the SCR and SDR will always be one, respectively. Combining the SCR and DSR together could tell how balanced the supply and demand of ES are. Lower and higher values of SDB will indicate more demand-limited and supply-limited realized ES, respectively. The zero value of SDB will represent a balanced condition of ES supply and demand, with fully consumed supply as well as fully satisfied demand (Fig. 3).

After using the framework to assess the actual use of multiple ES, multiple realized ES of the same SSA or SDA can be combined and analyzed together (Table 2). For example, the number of ES with AU values greater than zero can be considered the richness of ES in the region under evaluation. Different realized ES can also be grouped into supply-limited, demand-limited, and S-D balanced types based on their SDB values. The average SDB values of multiple realized ES provide people with information regarding the overall supply-demand balance condition in the region under evaluation (Table 2).

**Table 2**

An illustrative example of multiple ecosystem service type grouping based on the supply-demand balance index.

| | Supply Consumption Rate | Demand Satisfaction Rate | Supply-Demand Balance Index | Ecosystem Service Types |
|---|---|---|---|---|
| ES 1 | 1 | 1 | 0 | S-D Balanced |
| ES 2 | 1 | 0.9 | 0.1 | Supply-limited |
| ES 3 | 1 | 0.8 | 0.2 | Supply-limited |
| ES 4 | 1 | 0.6 | 0.4 | Supply-limited |
| ES 5 | 0.9 | 1 | -0.1 | Demand-limited |
| ES 6 | 0.8 | 1 | -0.2 | Demand-limited |
| ES 7 | 0.5 | 1 | -0.5 | Demand-limited |
| **Averages** | 0.89 | 0.90 | -0.01 | More demand-limited |

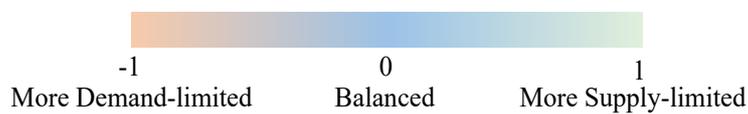

-1　　　　　　　　0　　　　　　　　1
More Demand-limited　　Balanced　　More Supply-limited

Note: only ecosystem services with actual use greater than zero should be considered.

## 3. Examples of Framework Application

To illustrate how the framework could work, we selected three additional ecosystem services that can be found in typical urban green parks and showed the different variables required for quantifying their actual use (Table 3). They are wild berry supply, pollination, and recreation, which represent a type of provisioning, regulating, and cultural services, respectively (Evans et al., 2022; Patrycja et al., 2022). Although the three are different types of ES, we aim to show that the process of using the SDFU framework to obtain the actual use is widely applicable to different ES types. Moreover, compared to services with more global scale demand (e.g., carbon sequestration), the three services all have a more localized demand in general and thus can be better examples showing the variable size of SDA identification.

*3.1 Wild Berry Supply*

First of all, we should identify the *Service Supply Area* (SSA) and *Service Demand Area* (SDA) of the wild berry supply service according to the SDFU framework. The SSA can be easily identified as the boundary of any urban green park under evaluation. Since urban green parks are most frequently visited by local residents, we deem that the SDA of the service could be neighboring communities of the urban park within walking distance, as these communities are the services where most needed (Berglihn and Gómez-Baggethun, 2021). In this case, the *Potential Supply* should be the total amount of edible and ripe wild berries in the park, which is the theoretical maximum capacity of ES supply. But not all of this capacity is certainly activated; rather, only the proportion of accessible berries would be qualified to be the *Max Actual SSA Supply*. Then, the export and import supply flows of wild berries, which need to be subtracted and added to the *Max Actual SSA Supply* to obtain the *Actual Supply*. The export and import flows could be the amount of berries carried out and into the SDA for consumption during evaluation, respectively. Moreover, the loss during supply-to-demand flow could be the number of wild berries lost before consumption, for instance, loss during transportation. Those possible losses should be subtracted to obtain the *Available Actual Supply* of wild berries.

On the other hand, the *Potential Demand* should be the amount of wild berries that can be consumed by the total resident population in the SDA. However, not all residents are certainly able and willing to eat wild berries from the urban park. Therefore, the *Max Actual SDA Demand* would only be the amount of wild berries that can be consumed by the proportion of residents who are able and willing to eat wild berries in the SDA during the evaluation. The export and import demand flows, which need to be subtracted and added to the *Max Actual SDA Demand* to obtain *Actual Demand*, could be the amount of wild berries that could be consumed by the residents, who are temporarily absent, and visitors, who are temporarily present, in the SDA during the evaluation, respectively. Lastly, the *Actual Use* of the service will equal the minimum quantity of wild berries of either the *Available Actual Supply* or *Actual Demand*.

*3.2 Pollination*

As with the wild berry supply service, the SSA of the pollination service is also the boundary of the urban green park. However, since pollination only becomes a final ecosystem service when natural pollination could benefit people directly, we deem agricultural plants that require natural pollination inside and neighbor the urban park within the maximum pollination radius (i.e., the active range of pollinators) as the SDA (Pellissier et al., 2012; Scheper et al., 2023). In this case, the *Potential Supply* would then be the capacity of agricultural plants that could be pollinated by the total population of pollinators in the park. But only a mature and healthy proportion of the total pollinator population can be deemed active pollinators. And only the number of agricultural plants that can be pollinated by these active populations can be considered the *Max Actual SSA Supply*. The export and import flows of pollination services could be the pollinators that move out and into the SDA during evaluation, respectively (which need to translate into the number of agricultural plants that can be pollinated by them to obtain the *Actual Supply*). The final *Available Actual Supply* could be *Actual Supply* minus the number of agricultural plants with failed pollination by active pollinators in the park for instance (Table 3).

The *Potential Demand* for the pollination service could be the total number of agricultural plants that require natural pollination in the SDA (Table 3). However, only the proportion at the stage of actually requiring pollination during the evaluation period could be considered the activated potential (i.e., the *Max Actual SDA Demand*). Since the growing plants at the stage of pollination are generally not moveable, there should not be export and import demand for the service. Thus, the *Max Actual SDA Demand* would just be the *Actual Demand*. Again, the *Actual Use* of pollination service will be constrained by the minimum number of agricultural plants in the *Available Actual Supply* or the *Actual Demand*.

*3.3 Recreation*

For the recreation service, the SSA for the recreation service would be the boundary of the urban green park again. The SDA can be considered the nearby communities of the urban park that are within walking distance because they would need the park's recreation service the most (Fischer et al., 2018; Perschke et al., 2023). In this case, the *Potential Supply* should be the visiting capacity (i.e., the maximum number of people allowed) of the whole area of the urban park. But only the activated capacity, the accessible and preferable area of the park in this case, during the evaluation could be the *Max Actual SSA Supply*. Since the recreation service in typical urban parks is usually an in situ service, which means that the service is provided and the benefit is realized in the same location (Fisher et al., 2009), there will be no export, import and supply-to-demand flows of the ES supply. Therefore, the *Max Actual SSA Supply* will be the same as the *Available Actual Supply*.

The *Potential Demand* for the recreation service could be the total resident population in the SDA. However, only the number of residents who are able and willing to recreate in the SDA during the evaluation should be considered the *Max Actual SDA Demand*. The export demand flow is the number of SDA residents who are able and willing to recreate but have their demand satisfied outside the SSA. In other words, these residents went to other places for recreation during the evaluation period. On the contrary, the import demand flow refers to the people, who recreate at the park but come from outside of the SDA, during the evaluation. Again, the *Actual Use* of recreation service in the park will be determined by the lower number of people of either the *Available Actual Supply* or the *Actual Demand*.

**Table 3**

Examples of the different variables required for quantifying the actual use of three ecosystem services can be found in typical urban green parks

| Ecosystem Services | Wild Berry Supply (Unit: amount of wild berry) | Pollination (Unit: number of agricultural plants) | Recreation (Unit: number of people) |
|---|---|---|---|
| *Service Supply Areas (SSA)* | Urban green park | Urban green park | Urban green park |
| *Service Demand Areas (SDA)* | Neighboring communities of the urban park within walking distance | Agricultural plants inside and neighboring the urban park within the max pollination range | Neighboring communities of the urban park within walking distance |
| *Potential Supply* | Total amount of edible and ripe wild berries in the park | Number of agricultural plants that can be pollinated by the total population of pollinators in the park | Visiting capacity (i.e., maximum number of people allowed) of the whole area of the urban park |
| *Max Actual SSA Supply* | Amount of accessible edible and ripe wild berries in the park during evaluation | Number of agricultural plants that can be pollinated by the amount of active (e.g., healthy and mature) pollinators in the park during evaluation | Visiting capacity of the accessible and preferable area of the park during evaluation |
| *Actual SSA Supply* | *Max Actual SSA Supply* minus the amount of wild berries carried out of the SDA for consumption during evaluation | *Max Actual SSA Supply* minus the number of agricultural plants that the pollinators could pollinate but move out of the SDA during evaluation | Equal to the *Max Actual SSA Supply* |
| *Actual Supply* | *Actual SSA Supply* plus the amount of wild berries carried into the SDA for consumption during evaluation | *Actual SSA Supply* plus the number of agricultural plants that were pollinated by the pollinators move into the SDA from outside during evaluation | Equal to the *Actual SSA Supply* |
| *Available Actual Supply* | *Actual Supply* minus the amount of wild berries lost before consumption (e.g., loss during transportation) | *Actual Supply* minus the number of agricultural plants with failed pollination by active pollinators in the park | Equal to the *Actual Supply* |
| *Potential demand* | Amount of wild berries can be consumed by the total resident population in the SDA of the park | Total number of agricultural plants grow in the SDA | Total resident population in the SDA |
| *Max Actual SDA Demand* | Amount of wild berries can be consumed by the proportion of residents who are able and willing to eat wild berries in the SDA during evaluation | Number of agricultural plants need natural pollination in the SDA during evaluation | Number of residents who are able and willing to recreate in the SDA during evaluation |
| *Actual SDA Demand* | *Max Actual SDA Demand* minus the amount of wild berries could be consumed by the residents who are temporarily absent in the SDA during evaluation | Equal to the *Max Actual SDA Demand* | *Max Actual SDA Demand* minus the number of residents who are able and willing to recreate but temporarily absent in the SDA during evaluation |
| *Actual Demand* | *Actual SDA Demand* plus the amount of wild berries could be consumed by the people who are temporarily present in the SDA during evaluation | Equal to the *Actual SDA Demand* | *Actual SDA Demand* plus the number of people who are able and willing to recreate and temporarily present in the SDA during evaluation |
| *Actual Use* | Minimum amount of wild berries of *Available Actual Supply* or *Actual Demand* | Minimum number of agricultural plants of *Available Actual Supply* or *Actual Demand* | Minimum number of people of *Available Actual Supply* or *Actual Demand* |

## 4. Discussion

*4.1 Scaling Features of Realized ES*

Since the provider of ecosystem services can range from individual plants to the whole planet, our proposed SDFU framework could also be used to assess the realized portion of their ES supply capacity at all ranges of spatial scales. One of the features when upscaling the SSA will be that the SDA would overlap more and more with the SSA. For instance, when we are assessing the actual use of ES provided by individual plants, the demand area for the services would probably not be the same as the supply area. However, when we are conducting global-scale studies and assessing the actual use of ES provided by the whole planet, the demand area would definitely overlap with the supply area.

Furthermore, another feature we could see when upscaling either SSA or SDA size is the increase in supply and demand quantity for ES. We presented an illustrative example of such Area-ES relationships between SSA-Supply and SDA-Demand in Figure 4. The slopes of SSA-Supply and SDA-Demand relationships represent the conversion rate of a land unit to the actual supply and demand of a specific ES, respectively. The real values will be determined by the quantity (e.g., vegetation coverage, urbanization rate, etc.) and quality (e.g., ecosystem structure, socio-economic development level, etc.) of the ES supplier and demander on the land unit. Although the slopes of the two relationships might not be the same, in theory, we could propose two sets of hypotheses that can be further tested by empirical data:

1. *For a given SSA and ES, there will be at least one size of SDA that can consume all the available actual supplies of the service in the SSA (i.e., an S-D balanced SDA).*
2. *For a given SDA and ES, there will be at least one size of SSA that can satisfy all the actual demands of the service in the SDA (i.e., an S-D balanced SSA).*

For example, in different scenarios when supply and demand are balanced, there will be cases where the area size of SSA is greater, equal, or smaller than the size of SDA (Fig. 4). The point of shifting depends on when the two lines interact, which is determined by the slopes of the two relationships (Fig. 4).

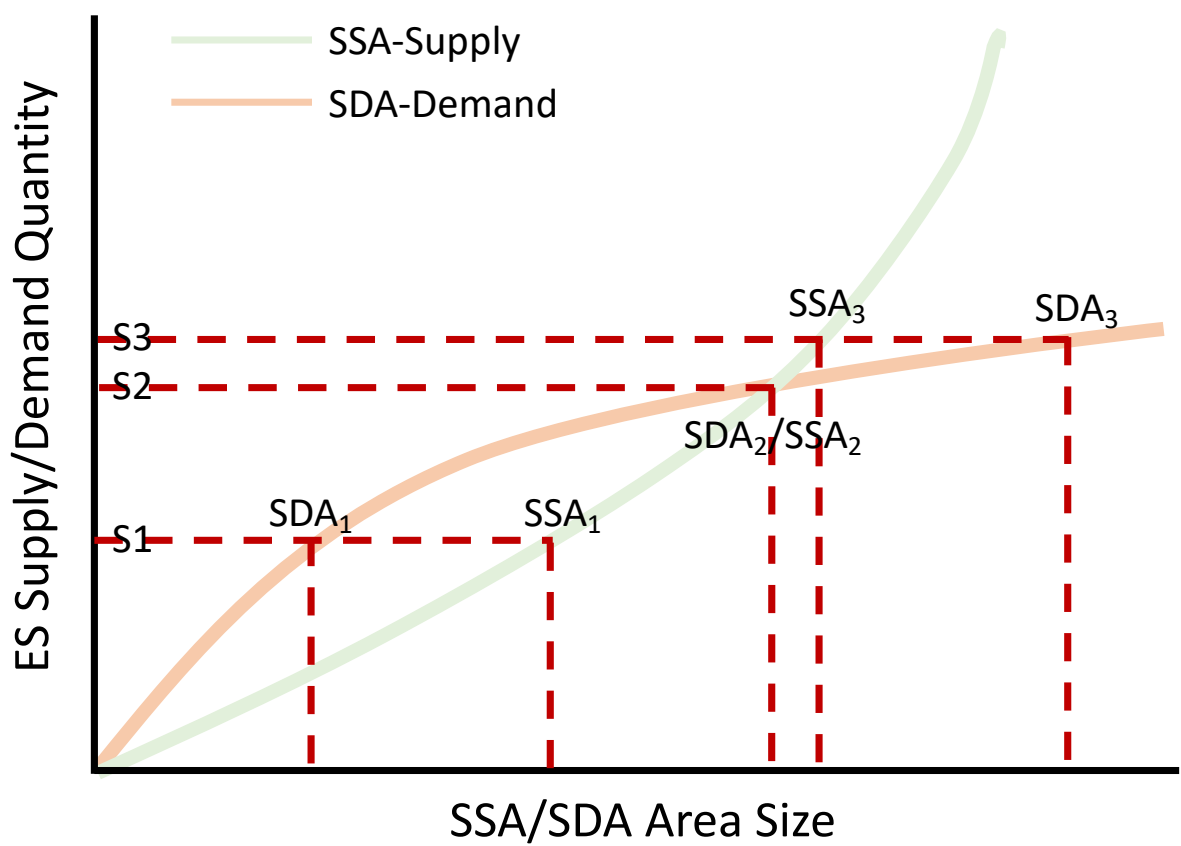

S-D Balanced Scenario 1 (S1): $SSA_1 > SDA_1$
S-D Balanced Scenario 2 (S2): $SSA_2 = SDA_2$
S-D Balanced Scenario 3 (S3): $SSA_3 < SDA_3$

**Fig. 4.** Schematic demonstrating of possible Area-ES relationships between SSA-Supply and SDA-Demand.

*4.2 Temporal Dynamics of Realized ES*

Since the actual use of ES is determined by all ES supply, demand, and flows, their changes in time could also cause temporal dynamics in realized ES. Multiple applications of the framework during the evaluation period might be necessary to catch such dynamic changes in realized ES. Many studies have pointed out the prevalence of longer-term changes (e.g., interannual dynamics) in ES supply and demand. For example, ecosystem degradation could reduce the capacity of the ecosystem to provide ES and thus lead to lower ES supply (Eddy et al., 2021), while urbanization could increase the total needs of socio-economic systems and thus lead to higher ES demand (Deng et al., 2021) (Fig. 5).

Nevertheless, we want to emphasize that the shorter-term temporal changes in ES supply and demand could also cause important dynamic patterns in realized ES. For instance, the seasonal change from winter (or dormant season) to summer (or growing season) could also lead to an increase in ES supply (Körner et al., 2023) (Fig. 5). Even temporal changes at diurnal scales, such as the shift from daytime to nighttime, could lead to a decrease in ES demand (Liang et al., 2020) (Fig. 5). Therefore, to better understand the temporal patterns of realized ES, we want to call for more studies on the dynamic changes in all ES supply, demand, and flows, especially at the shorter-time scale.

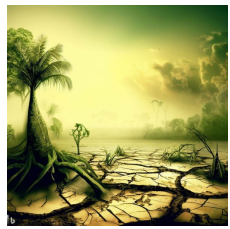 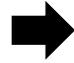 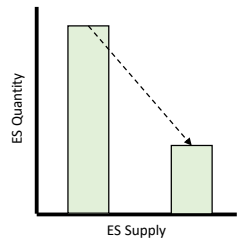 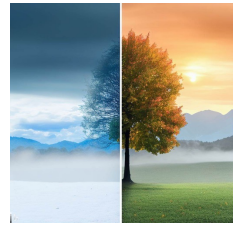 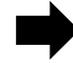 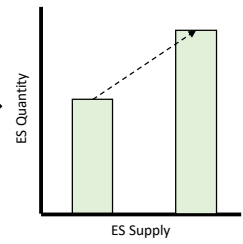
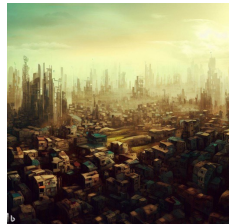 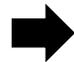 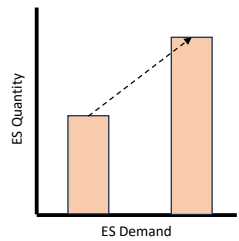 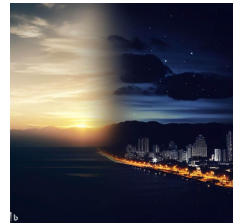 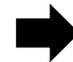 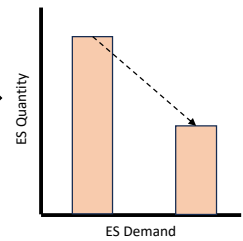

**Fig. 5.** Examples of longer-term and shorter-term temporal dynamics in ecosystem service supply and demand changes. Images of examples are drawn by DALL-E.

*4.3 Spatial Characteristics of Realized ES*

The first spatial characteristic we want to discuss is the spatial relationship between SSA and SDA (Fisher et al., 2009; Vrebos et al., 2015). These spatial relationships would determine if flows need to be considered when evaluating realized ES or not. For example, the ES supply import, export, and supply-to-demand flows can only be found for ES with omni-directional and directional characteristics but not for in situ services (Vrebos et al., 2015). If the SSA and SDA are in the same location (i.e., in situ), then the *Available Actual Supply* will just equal the *Max Actual SSA Supply* (as in our example of the recreation service supply in Table 3). The key to understanding whether a service is in situ or not depends on whether the service carrier is mobile or immobile (Bagstad et al., 2013). Similarly, the import and export flows of ES demand can only be found for mobile service demand agents (e.g., people, greenhouse gases, pests), but not for immobile ones (e.g., crops needing pollination). If the demander of the service is not movable, then the *Actual Demand* will just be the same as the *Max Actual SDA Demand* (as in our example of the pollination service demand in Table 3.

In addition, many studies have now noted another important spatial characteristic of ecosystem service distribution, which is spatial heterogeneity. Similar to the temporal dynamics, the different spatial patterns of the realized ES also originated from the spatial heterogeneity of ES supply, demand, and flows (Lin et al., 2022; Perschke et al., 2023). Recent studies have shown that the hotspots and coldspots for the supply, demand, and flows of various services, such as stormwater regulation (Goldenberg et al., 2017), aesthetic experiences (Egarter Vigl et al., 2017), nature-based recreation (Aziz, 2023) and urban thermoregulation (Xu et al., 2023), could all be highly variable in space. These variations would certainly lead to highly heterogeneous spatial distributions in the actual use of ES.

*4.4 Further Questions*

Last but not least, we want to stress that the proposed SDFU framework is far from an end but a starting point for better analyzing realized ES. Many challenges and unsolved questions remain for further investigations to improve our understanding of the actual use of ES. Here, we would like to list some critical questions whose answers might be crucial for advancing the study of realized ES:

1. **How to accurately quantify the different components required to calculate the actual use of ES?** This question is the most paramount to the proposed framework since it relates to its applicability in the real world. It must be acknowledged that some of the measurements are very hard to acquire, such as the export and import flows of both ES supply and demand (Li and Wang, 2023; Wang et al., 2023). However, with more careful experimental designs (e.g., more extensive sampling across boundaries of SSA and SDA) and advances in data-collecting technologies (e.g., increasing affordable high-throughput recording hardware and automated technologies for fast and accurate ecosystem monitoring) (Besson et al., 2022), we believe that more and more high-quality data would be available in the future.

2. **When do supply-limited and demand-limited ES shift?** There might be two approaches to addressing this question. When the size of SSA and/or SDA is fixed (e.g., the administrative boundary of a city) and data across multiple times (e.g., panel data) are available, then the temporal change patterns of the actual use of ES can be delineated to answer the question. On the other hand, when there is only a single opportunity for evaluation (e.g., a snapshot of data) and the size of SSA or SDA is not fixed, the answer to the question would become to depict the Area-ES relationship (e.g., Fig. 4) to find the size of S-D balanced SSA or SDA.

3. **What are the spatial-temporal dynamics in the actual use of ES?** As we discussed earlier, the actual use of ES could be both temporally dynamic and spatially heterogeneous. Therefore, quantifying and comparing the spatiotemporal changing patterns of various realized ES would be essential for understanding their characteristics. This might require spatially explicit mapping of the actual use of various ES (obtained through the application of the SDFU framework) over multiple time periods to show the dynamic changes of different realized ES both spatially and temporally at the same time (Boesing et al., 2020).

4. **What are the influencing factors in the actual use of ES?** The SDFU framework provides a full list of the variables required for assessing the actual use of ES (Table 1). These variables are all direct determinants of the actual use of ES. However, each of the variables is also influenced by a range of ecological, environmental, social, and economic factors. Some of the factors might also influence multiple variables at the same time (e.g., climate conditions, such as warming temperatures, can affect both supply and demand of ES, such as water supply) (Abdulla et al., 2009). Therefore, it is crucial to consider the full scope of the influencing factors and their interactions in analyses to accurately understand the effects of the influencing factors and predict their impacts on the actual use of ES.

5. **What are the roles of local vs. external supply and demand in the S-D balance?** A more specific influencing factor in balancing the actual use of ES will be how much supply is provided as well as demand is needed locally vs. externally. A local or external dominant region could mean completely different management strategies for achieving S-D balance in the actual use of ES. Answering this question requires accurate quantification of all of the export and import flows of ES supply and demand. Coefficients, such as the ratio of *Actual Use* to *Actual SSA Supply* and *Actual SDA Demand*, could also be developed to indicate the role local supply and demand play in S-D balance.

6. **How to manage to enhance the sustainable actual use of multiple ES?** This question comes from the perspective of management and policy-making. By making management plans and policies that can sustainably improve the actual use of different ES, people would be able not only to better achieve global visions for human society like the Sustainable Development Goals (Yin et al., 2021), but also improve the conditions of natural ecosystems and contribute to fulfilling the targets, such as the Kunming-Montreal Global Biodiversity Framework (Shen et al., 2023). However, such plans and policies must be based on scientific evidence obtained through answering questions like the aforementioned.

**5. Conclusions**

In this study, we proposed a Supply-Demand-Flow-Use (SDFU) framework to analyze the realized ecosystem services (ES) that are directly linked to human well-being. The framework integrates the supply, demand, flow, and use of ES and differentiates them into various aspects to better understand the actual use of ES. We applied the framework to three examples of ES that can be found in typical urban green parks: wild berry supply, pollination, and recreation. We showed how the framework could assess the actual use of ES and identify the supply-limited, demand-limited, and supply-demand-balanced types of realized ES. We also discussed the scaling features, temporal dynamics, and spatial characteristics of realized ES, as well as some crucial questions for future research.

We believe the application of the SDFU framework can provide a comprehensive and systematic way to assess the actual use of ES and inform management and policy-making for the sustainable use of nature's benefits. For example, trade-offs and synergies among multiple realized ES can be evaluated, and their allocation and distribution could then be optimized. However, we also acknowledge that the framework still faces multiple challenges, such as data availability, quantification methods, and scaling issues. More empirical evidence and applications are needed to improve the applicability and robustness of the framework and advance the study of realized ES. We hope that our study will stimulate more future research on realized ES and contribute to a better understanding of their roles in enhancing human well-being and conserving natural ecosystems.

**Acknowledgments**

We thank the Shandong Provincial Natural Science Foundation (ZR2022QC253) for funding this research.

# References

Abdulla, F., Eshtawi, T., Assaf, H., 2009. Assessment of the Impact of Potential Climate Change on the Water Balance of a Semi-arid Watershed. Water Resour. Manag. 23, 2051–2068. https://doi.org/10.1007/s11269-008-9369-y

Aziz, T., 2023. Terrestrial protected areas: Understanding the spatial variation of potential and realized ecosystem services. J. Environ. Manage. 326, 116803. https://doi.org/10.1016/j.jenvman.2022.116803

Aziz, T., Shah, M.U., 2019. Realized Ecosystem Services: Using Stakeholder Theory for Policy Development 2.

Bagstad, K.J., Johnson, G.W., Voigt, B., Villa, F., 2013. Spatial dynamics of ecosystem service flows: A comprehensive approach to quantifying actual services. Ecosyst. Serv. 4, 117–125. https://doi.org/10.1016/j.ecoser.2012.07.012

Berglihn, E.C., Gómez-Baggethun, E., 2021. Ecosystem services from urban forests: The case of Oslomarka, Norway. Ecosyst. Serv. 51, 101358. https://doi.org/10.1016/j.ecoser.2021.101358

Besson, M., Alison, J., Bjerge, K., Gorochowski, T.E., Høye, T.T., Jucker, T., Mann, H.M.R., Clements, C.F., 2022. Towards the fully automated monitoring of ecological communities. Ecol. Lett. 25, 2753–2775. https://doi.org/10.1111/ele.14123

Boesing, A.L., Prist, P.R., Barreto, J., Hohlenwerger, C., Maron, M., Rhodes, J.R., Romanini, E., Tambosi, L.R., Vidal, M., Metzger, J.P., 2020. Ecosystem services at risk: integrating spatiotemporal dynamics of supply and demand to promote long-term provision. One Earth 3, 704–713. https://doi.org/10.1016/j.oneear.2020.11.003

Brauman, K.A., Garibaldi, L.A., Polasky, S., Aumeeruddy-Thomas, Y., Brancalion, P.H.S., DeClerck, F., Jacob, U., Mastrangelo, M.E., Nkongolo, N.V., Palang, H., Pérez-Méndez, N., Shannon, L.J., Shrestha, U.B., Strombom, E., Verma, M., 2020. Global trends in nature's contributions to people. Proc. Natl. Acad. Sci. 117, 32799–32805. https://doi.org/10.1073/pnas.2010473117

Burkhard, B., Kandziora, M., Hou, Y., Müller, F., 2014. Ecosystem service potentials, flows and demands-concepts for spatial localisation, indication and quantification. Landsc. Online 34, 1–32. https://doi.org/10.3097/LO.201434

Costanza, R., d'Arge, R., de Groot, R., Farber, S., Grasso, M., Hannon, B., Limburg, K., Naeem, S., O'Neill, R.V., Paruelo, J., Raskin, R.G., Sutton, P., van den Belt, M.,

1997. The value of the world's ecosystem services and natural capital. Nature 387, 253–260. https://doi.org/10.1038/387253a0

Costanza, R., de Groot, R., Braat, L., Kubiszewski, I., Fioramonti, L., Sutton, P., Farber, S., Grasso, M., 2017. Twenty years of ecosystem services: How far have we come and how far do we still need to go? Ecosyst. Serv. 28, 1–16. https://doi.org/10.1016/j.ecoser.2017.09.008

Dangles, O., Casas, J., 2019. Ecosystem services provided by insects for achieving sustainable development goals. Ecosyst. Serv. 35, 109–115. https://doi.org/10.1016/j.ecoser.2018.12.002

Deng, C., Liu, J., Liu, Y., Li, Z., Nie, X., Hu, X., Wang, L., Zhang, Y., Zhang, G., Zhu, D., Xiao, L., 2021. Spatiotemporal dislocation of urbanization and ecological construction increased the ecosystem service supply and demand imbalance. J. Environ. Manage. 288, 112478. https://doi.org/10.1016/j.jenvman.2021.112478

Eddy, T.D., Lam, V.W.Y., Reygondeau, G., Cisneros-Montemayor, A.M., Greer, K., Palomares, M.L.D., Bruno, J.F., Ota, Y., Cheung, W.W.L., 2021. Global decline in capacity of coral reefs to provide ecosystem services. One Earth 4, 1278–1285. https://doi.org/10.1016/j.oneear.2021.08.016

Egarter Vigl, L., Depellegrin, D., Pereira, P., De Groot, R., Tappeiner, U., 2017. Mapping the ecosystem service delivery chain: Capacity, flow, and demand pertaining to aesthetic experiences in mountain landscapes. Sci. Total Environ. 574, 422–436. https://doi.org/10.1016/j.scitotenv.2016.08.209

Evans, D.L., Falagán, N., Hardman, C.A., Kourmpetli, S., Liu, L., Mead, B.R., Davies, J.A.C., 2022. Ecosystem service delivery by urban agriculture and green infrastructure – a systematic review. Ecosyst. Serv. 54, 101405. https://doi.org/10.1016/j.ecoser.2022.101405

Fischer, L.K., Honold, J., Botzat, A., Brinkmeyer, D., Cvejić, R., Delshammar, T., Elands, B., Haase, D., Kabisch, N., Karle, S.J., Lafortezza, R., Nastran, M., Nielsen, A.B., Van Der Jagt, A.P., Vierikko, K., Kowarik, I., 2018. Recreational ecosystem services in European cities: Sociocultural and geographical contexts matter for park use. Ecosyst. Serv. 31, 455–467. https://doi.org/10.1016/j.ecoser.2018.01.015

Fisher, B., Turner, R.K., Morling, P., 2009. Defining and classifying ecosystem services for decision making. Ecol. Econ. 68, 643–653. https://doi.org/10.1016/j.ecolecon.2008.09.014

Goldenberg, R., Kalantari, Z., Cvetkovic, V., Mörtberg, U., Deal, B., Destouni, G., 2017. Distinction, quantification and mapping of potential and realized supply-demand of flow-dependent ecosystem services. Sci. Total Environ. 593–594, 599–609. https://doi.org/10.1016/j.scitotenv.2017.03.130

Haines-Young, R., Potschin, M., 2010. The links between biodiversity, ecosystem services and human well-being, in: Raffaelli, D.G., Frid, C.L.J. (Eds.), Ecosystem Ecology. Cambridge University Press, Cambridge, pp. 110–139. https://doi.org/10.1017/CBO9780511750458.007

Jones, L., Norton, L., Austin, Z., Browne, A.L., Donovan, D., Emmett, B.A., Grabowski, Z.J., Howard, D.C., Jones, J.P.G., Kenter, J.O., Manley, W., Morris, C., Robinson, D.A., Short, C., Siriwardena, G.M., Stevens, C.J., Storkey, J., Waters, R.D., Willis, G.F., 2016. Stocks and flows of natural and human-derived capital in ecosystem services. Land Use Policy 52, 151–162. https://doi.org/10.1016/j.landusepol.2015.12.014

Kleemann, J., Schröter, M., Bagstad, K.J., Kuhlicke, C., Kastner, T., Fridman, D., Schulp, C.J.E., Wolff, S., Martínez-López, J., Koellner, T., Arnhold, S., Martín-López, B., Marques, A., Lopez-Hoffman, L., Liu, J., Kissinger, M., Guerra, C.A., Bonn, A., 2020. Quantifying interregional flows of multiple ecosystem services – A case study for Germany. Glob. Environ. Change 61, 102051. https://doi.org/10.1016/j.gloenvcha.2020.102051

Körner, C., Möhl, P., Hiltbrunner, E., 2023. Four ways to define the growing season. Ecol. Lett. ele.14260. https://doi.org/10.1111/ele.14260

Kremen, C., Chaplin-Kramer, R., 2007. Insects as providers of ecosystem services: crop pollination and pest control. Insect Conserv. Biol., CABI Books 349–382. https://doi.org/10.1079/9781845932541.0349

La Notte, A., Vallecillo, S., Marques, A., Maes, J., 2019. Beyond the economic boundaries to account for ecosystem services. Ecosyst. Serv. 35, 116–129. https://doi.org/10.1016/j.ecoser.2018.12.007

Li, J., Wang, Y., 2023. Ecosystem services assessment from capacity to flow: A review. Trans. Earth Environ. Sustain. 1, 80–93. https://doi.org/10.1177/2754124X221141991

Liang, Z., Wang, Y., Huang, J., Wei, F., Wu, S., Shen, J., Sun, F., Li, S., 2020. Seasonal and Diurnal Variations in the Relationships between Urban Form and the Urban Heat Island Effect. Energies 13, 5909. https://doi.org/10.3390/en13225909

Lin, Y., Zhang, M., Gan, M., Huang, L., Zhu, C., Zheng, Q., You, S., Ye, Z., Shahtahmassebi, A., Li, Y., Deng, J., Zhang, J., Zhang, L., Wang, K., 2022. Fine identification of the supply–demand mismatches and matches of urban green space ecosystem services with a spatial filtering tool. J. Clean. Prod. 336, 130404. https://doi.org/10.1016/j.jclepro.2022.130404

Liu, J., 2023. Leveraging the metacoupling framework for sustainability science and global sustainable development. Natl. Sci. Rev. 10, nwad090. https://doi.org/10.1093/nsr/nwad090

Liu, Y., Fu, B., Wang, S., Rhodes, J.R., Li, Y., Zhao, W., Li, C., Zhou, S., Wang, C., 2023. Global assessment of nature's contributions to people. Sci. Bull. 68, 424–435. https://doi.org/10.1016/j.scib.2023.01.027

Patrycja, P., Krzysztof, M., Marcin, M., Adam, I., Piotr, M., 2022. Ranking ecosystem services delivered by trees in urban and rural areas. Ambio 51, 2043–2057. https://doi.org/10.1007/s13280-022-01722-2

Pellissier, V., Muratet, A., Verfaillie, F., Machon, N., 2012. Pollination success of Lotus corniculatus (L.) in an urban context. Acta Oecologica 39, 94–100. https://doi.org/10.1016/j.actao.2012.01.008

Peng, J., Xia, P., Liu, Y., Xu, Z., Zheng, H., Lan, T., Yu, S., 2023. Ecosystem services research: From golden era to next crossing. Trans. Earth Environ. Sustain. 1, 9–19. https://doi.org/10.1177/2754124X231165935

Perschke, M.J., Harris, L.R., Sink, K.J., Lombard, A.T., 2023. Using ecological infrastructure to comprehensively map ecosystem service demand, flow and capacity for spatial assessment and planning. Ecosyst. Serv. 62, 101536. https://doi.org/10.1016/j.ecoser.2023.101536

Rasmussen, L.V., Mertz, O., Christensen, A.E., Danielsen, F., Dawson, N., Xaydongvanh, P., 2016. A combination of methods needed to assess the actual use of provisioning ecosystem services. Ecosyst. Serv. 17, 75–86.
https://doi.org/10.1016/j.ecoser.2015.11.005

Scheper, J., Badenhausser, I., Kantelhardt, J., Kirchweger, S., Bartomeus, I., Bretagnolle, V., Clough, Y., Gross, N., Raemakers, I., Vilà, M., Zaragoza-Trello, C., Kleijn, D., 2023. Biodiversity and pollination benefits trade off against profit in an intensive farming system. Proc. Natl. Acad. Sci. 120, e2212124120. https://doi.org/10.1073/pnas.2212124120

Schröter, M., Koellner, T., Alkemade, R., Arnhold, S., Bagstad, K.J., Erb, K.-H., Frank, K., Kastner, T., Kissinger, M., Liu, J., López-Hoffman, L., Maes, J., Marques, A., Martín-López, B., Meyer, C., Schulp, C.J.E., Thober, J., Wolff, S., Bonn, A., 2018. Interregional flows of ecosystem services: Concepts, typology and four cases. Ecosyst. Serv. 31, 231–241. https://doi.org/10.1016/j.ecoser.2018.02.003

Serna-Chavez, H.M., Schulp, C.J.E., Van Bodegom, P.M., Bouten, W., Verburg, P.H., Davidson, M.D., 2014. A quantitative framework for assessing spatial flows of ecosystem services. Ecol. Indic. 39, 24–33. https://doi.org/10.1016/j.ecolind.2013.11.024

Shen, X., Liu, M., Hanson, J.O., Wang, J., Locke, H., Watson, J.E.M., Ellis, E.C., Li, S., Ma, K., 2023. Countries' differentiated responsibilities to fulfill area-based conservation targets of the Kunming-Montreal Global Biodiversity Framework. One Earth 6, 548–559. https://doi.org/10.1016/j.oneear.2023.04.007

United Nations, 2021. System of Environmental-Economic Accounting--Ecosystem Accounting (SEEA EA).

Vandermeer, J., Perfecto, I., Philpott, S., 2010. Ecological Complexity and Pest Control in Organic Coffee Production: Uncovering an Autonomous Ecosystem Service. BioScience 60, 527–537. https://doi.org/10.1525/bio.2010.60.7.8

Villamagna, A.M., Angermeier, P.L., Bennett, E.M., 2013. Capacity, pressure, demand, and flow: A conceptual framework for analyzing ecosystem service provision and delivery. Ecol. Complex. 15, 114–121. https://doi.org/10.1016/j.ecocom.2013.07.004

Vrebos, D., Staes, J., Vandenbroucke, T., D׳Haeyer, T., Johnston, R., Muhumuza, M., Kasabeke, C., Meire, P., 2015. Mapping ecosystem service flows with land cover scoring maps for data-scarce regions. Ecosyst. Serv. 13, 28–40. https://doi.org/10.1016/j.ecoser.2014.11.005

Wang, L., Wu, T., Zheng, H., Li, R., Hu, X., Ouyang, Z., 2023. A comprehensive framework for quantifying ecosystem service flow focusing on social-ecological processes. Trans. Earth Environ. Sustain. 1, 20–34. https://doi.org/10.1177/2754124X231164797

Wang, L., Zheng, H., Chen, Y., Ouyang, Z., Hu, X., 2022. Systematic review of ecosystem services flow measurement: Main concepts, methods, applications and future directions. Ecosyst. Serv. 58, 101479. https://doi.org/10.1016/j.ecoser.2022.101479

Wei, H., Fan, W., Wang, X., Lu, N., Dong, X., Zhao, Yanan, Ya, X., Zhao, Yifei, 2017. Integrating supply and social demand in ecosystem services assessment: A review. Ecosyst. Serv. 25, 15–27. https://doi.org/10.1016/j.ecoser.2017.03.017

Wong, C.P., Jiang, B., Kinzig, A.P., Lee, K.N., Ouyang, Z., 2015. Linking ecosystem characteristics to final ecosystem services for public policy. Ecol. Lett. 18, 108–118. https://doi.org/10.1111/ele.12389

Xu, L., Hang, Y., Jin, F., 2023. Evaluating the supply-demand relationship for urban green parks in Beijing from an ecosystem service flow perspective. Urban For. Urban Green. 85, 127974. https://doi.org/10.1016/j.ufug.2023.127974

Yin, C., Zhao, W., Cherubini, F., Pereira, P., 2021. Integrate ecosystem services into socio-economic development to enhance achievement of sustainable development goals in the post-pandemic era. Geogr. Sustain. 2, 68–73. https://doi.org/10.1016/j.geosus.2021.03.002

Zeng, J., Cui, X., Chen, W., Yao, X., 2023. Ecological management zoning based on the supply-demand relationship of ecosystem services in China. Appl. Geogr. 155, 102959. https://doi.org/10.1016/j.apgeog.2023.102959